\documentclass[twocolumn,aps,floatfix,showpacs,superscriptaddress]{revtex4}
\usepackage{amsmath,amssymb,eucal,graphicx}

\begin{document}

\title{Probing Non-Integer Dimensions}

\author{P.~L.~Krapivsky} \affiliation{Center for Polymer Studies and Department of Physics, Boston
  University, Boston, Massachusetts 02215, USA}
\author{S.~Redner} \affiliation{Center for Polymer Studies and Department of Physics, Boston
  University, Boston, Massachusetts 02215, USA}

\begin{abstract} 
  We show that two-dimensional convection-diffusion problems with a radial
  sink or source at the origin may be recast as a pure diffusion problem in a
  fictitious space in which the spatial dimension is continuously-tunable
  with the P\'eclet number. This formulation allows us to probe
  various diffusion-controlled processes in non-integer dimensions.
\end{abstract}

\maketitle

\section{Introduction}

Consider a system of non-interacting passive particles that undergo a
combination of diffusion and convection that is driven by a steady velocity
field ${\bf v}\equiv {\bf v}({\bf r})$. The particle density $c\equiv c({\bf
  r},t)$ obeys the equation of motion
\begin{equation}
\label{general}
\frac{\partial c}{\partial t}+\left({\bf v}\cdot\nabla\right)c
=D\nabla^2 c,
\end{equation}
where $D$ the diffusion coefficient.  If the velocity field is radial, ${\bf
  v}=v(r)\,\hat r$, and the initial conditions are radially symmetric, the
density satisfies
\begin{equation}
\label{radial}
\frac{\partial c}{\partial t}+v\frac{\partial c}{\partial r}
=D\left(\frac{\partial^2 c}{\partial r^2}
+\frac{d-1}{r}\,\frac{\partial c}{\partial r}\right).
\end{equation}

Suppose that the velocity field is inversely proportional to the radial
distance, $v=\frac{Q}{2\pi r}$.  Such a divergenceless field is natural in
two dimensions and is generated by a point sink (or source) of strength $Q$
in an incompressible fluid.  This flow field has the remarkable property that
the convection term can be absorbed into the diffusion operator by an
appropriate shift of the spatial dimension. Indeed, the density satisfies
\begin{equation}
\label{diff}
\frac{\partial c}{\partial t}
=D\left(\frac{\partial^2 c}{\partial r^2}
+\frac{d_{\rm eff}-1}{r}\,\frac{\partial c}{\partial r}\right),
\end{equation}
where the effective spatial dimension is given by
\begin{equation}
\label{dim}
d_{\rm eff}=d-\frac{Q}{2\pi D}=2-2\lambda, \quad \lambda=\frac{Q}{4\pi D}.
\end{equation}
Here $\lambda$ is the P\'eclet number; the numerical factor $(4\pi)^{-1}$ is
chosen to simplify formulae that follow.

We can thus interpret convection-diffusion in two dimensions with a radial
velocity $v\propto r^{-1}$ as isotropic diffusion in a space with variable
effective dimension.  The motivation for developing this connection is that
there are many problems in many-body physics for which the spatial dimension
is an important determining factor in the phenomenology.  In
diffusion-controlled reactions, for example, a large body of work has
uncovered the general feature that when the spatial dimension $d$ exceeds a
critical value $d_c$ (that depends on the specifics of the reaction), the
reaction kinetics has a mean-field character, a situation where any pair of
reactants is equally likely to react. Conversely, when the spatial dimension
$d<d_c$, fluctuation-dominated phenomena arise, such as anomalously slow
kinetics and non-trivial spatial organization of reactants, see, {\it e.g.},
Refs.~\cite{ov,redner,oshanin,priv,dani,fpp} for a review.  The radial flow system offers a
potentially attractive way to study the full range of behavior between the
disparate mean-field and fluctuation-controlled regimes simply by tuning the
flow velocity.

In this work, we therefore exploit the connection between
convection-diffusion in two dimensions and pure diffusion in a space with a
tunable effective dimension.  By this equivalence we can probe
diffusion-controlled processes in arbitrary dimension by varying the P\'eclet
number.  In the next section we emphasize some subtleties associated with
this mapping.  In the following sections, we study simple
diffusion-controlled reactions in general dimensions by exploiting the
mapping to radial flow in two dimensions.  We first consider the influence of
an absorbing trap on the concentration profile in Sect.~\ref{trap}.  In
particular, we analyze the minimal separation between the particles and the
trap.  In Sect.~\ref{coag} we consider two simple reactive systems, namely,
irreversible coalescence and irreversible annihilation.  

\section{Spreading of a ring}
\label{simple}

As a warm-up exercise, we first consider the spread of an initial density
profile that is concentrated on a ring of radius $R$,
\begin{equation}
\label{init}
c(r,t=0)=\frac{1}{2\pi R}\,\delta(r-R).
\end{equation}
This example illuminates some pitfalls of the mapping convection-diffusion in
two dimensions onto a pure diffusion problem and shows that the size of the
ring is asymptotically relevant only for sink flows.

It is possible to find the spread of the initial ring by solving the
convection-diffusion equation in two dimensions subject to the initial
condition \eqref{init}.  The solution can be written in terms of Bessel
functions in the Laplace domain or as an infinite series in the time domain.
However, physical insight about the solution is more easily obtained from its
asymptotic behavior.  To determine the asymptotics, we make use of the
mapping \eqref{diff} and express the original problem as a purely diffusive
system and then use the well-known scaling solution of the latter equation to
describe the spread from a point mass at the origin.

There is a subtlety in specifying the initial condition.  The
initial two-dimensional ring distribution \eqref{init} is normalized to unit
``mass''
\begin{equation*}
\mathcal{M}(t=0)=\int_0^\infty c(r,t=0)\,2\pi r\,dr=1\,.
\end{equation*}
In the effective $d$-dimensional space, however, the
mass $M_{\rm eff}$ is different, 
\begin{equation}
\label{mass}
M_{\rm eff}=\int_0^\infty c(r,t=0)\,\Omega_d r^{d-1}dr
=\frac{(\pi R^2)^{-\lambda}}{\Gamma(1-\lambda)}
\end{equation}
Here $\Omega_d=2\pi^{d/2}/\Gamma(d/2)$ is the surface area of unit sphere in
$d$ dimensions.  Asymptotically, the concentration profile should depend on
the total effective mass $M_{\rm eff}$, while the initial size $R$ should
become irrelevant.  Thus one might anticipate that the density approaches
\begin{equation}
\label{main}
c(r,t)\to \frac{M_{\rm eff}}{(4\pi Dt)^{d/2}}\,\,\exp\left(-\frac{r^2}{4Dt}\right).
\end{equation}
in the long-time limit. 

Interestingly, Eq.~\eqref{main} is correct only for sink flows.  An
inconsistency in the ansatz \eqref{main} can immediately be seen by computing
the total mass, $\mathcal{M}(t)=\int c(r,t)2\pi r\,dr$, in the {\em physical}
two-dimensional space.  Using \eqref{main} we obtain
\begin{equation}
\label{Pt}
\mathcal{M}(t)\to \frac{1}{\Gamma(1-\lambda)}\,\left(\frac{4Dt}{R^2}\right)^\lambda
\end{equation}
in the long-time limit.  Equation \eqref{Pt} looks reasonable for sink
flows---the sink causes particles to disappear which is reflected by the
decay of $\mathcal{M}(t)$.  On the other hand, for source flows ($\lambda>0$)
equation \eqref{Pt} is invalid---the prediction that the total mass diverges
as $t^{\lambda}$ is clearly non-sensical.

To resolve this puzzle, it is useful to recall the derivation of
Eq.~\eqref{main} to see why it does not apply to source flows. The fundamental
solution \eqref{main} may be found by noting that the diffusion equation is
invariant under the transformation $r\to ar, t\to a^2t$.  Hence the scaling
variable $\eta=r^2/4Dt$ remains invariant under this scale transformation,
suggesting that the fundamental solution has the form $c(r,t)=t^{-b}F(\eta)$.
Then mass conservation $\int d^dr\, c(r,t)=M={\rm const.}$ yields $b=d/2$.
Finally, by substituting $c(r,t)=t^{-d/2}F(\eta)$ into the diffusion equation
and solving the resulting ordinary differential equation we find
$F=e^{-\eta}$ and thus recover Eq.~\eqref{main}.  

For the radial flow problem, however, mass conservation arises in the
physical two-dimensional space, rather than in the effective $d$-dimensional
space.  Therefore, $b=d_{\rm phys}/2=1$; that is, the asymptotic solution
should read
\begin{equation}
\label{crt}
c(r,t)\to \frac{1}{4\pi Dt}\,F(\eta), \quad \eta=\frac{r^2}{4Dt},
\end{equation}
so that the conservation law
\begin{equation}
\label{cons}
\mathcal{M}(t)=\int_0^\infty c(r,t)\,2\pi r\,dr=\int_0^\infty F(\eta)\,d\eta=1
\end{equation}
indeed holds.  By substituting (\ref{crt}) into the governing diffusion equation
(\ref{diff}) we find
\begin{equation}
\label{F}
\eta F''+(1-\lambda+\eta)F'+F=0,
\end{equation}
which is solved to yield 
\begin{equation}
\label{Feta}
F(\eta)=\frac{\eta^\lambda}{\Gamma(1+\lambda)}\,e^{-\eta}.
\end{equation}

Thus for source flows, the total mass remains constant, $\mathcal{M}(t)\equiv
1$, and the density is
\begin{eqnarray*}
c(r,t)\to \frac{1}{ \Gamma(1+\lambda)}\,\frac{1}{4\pi Dt}\,
\left(\frac{r^2}{4Dt}\right)^\lambda\,\exp\left(-\frac{r^2}{4Dt}\right).
\end{eqnarray*}
In this case, the system ``forgets'' the initial conditions.  

Conversely, our hand-waving argument is correct for sink flows as can be
verified by explicit solution of the full convection-diffusion equation
\cite{fpp}.  Therefore the total mass decays according to Eq.~\eqref{Pt} and
the density is given by \eqref{main}, or equivalently
\begin{eqnarray*}
c(r,t)\to \frac{1}{\Gamma(1-\lambda)}\,\frac{1}{4\pi Dt}\,
\left(\frac{4Dt}{R^2}\right)^\lambda\,\exp\left(-\frac{r^2}{4Dt}\right).
\end{eqnarray*}
The system ``remembers'' initial conditions as it is reflected by
the appearance of $R$ in the above asymptotic expressions. 

\section{Distance to the Closest Particle}
\label{trap}

As an illustration of the utility of the equivalence between
convection-diffusion in two dimensions and pure diffusion in general
dimensions, we examine here the following question: What is the typical
distance from the boundary of the absorbing trap to the closest particle?  
Suppose that the trap is a disc of radius $R$ centered at the origin, so that 
particles freely diffuse for $r>R$ and get absorbed at $r=R$.
We assume that initial density is uniform, $c|_{t=0}=c_0$. We anticipate that
the typical separation $r_{\rm min}$ between the trap and the
closest surviving particle exhibits the following behaviors:
\begin{itemize}
\item For sink flows, $r_{\rm min}$ is finite.
\item For source flows, $r_{\rm min}$ diverges with time.
\end{itemize}
We now confirm these expectations and establish the precise asymptotic form
of the minimal distance $r_{\rm min}$ by employing a simple quasi-static
approach \cite{rb,k}.  The minimal distance can also be found by more formal
methods \cite{WHK,HLKW}.

For sink flows, the concentration approaches a time-independent profile
$c_\infty(r)$ that is found by solving the Laplace equation subject to the
absorbing boundary condition, $c_\infty|_{r=R}=0$.  This solution is
\begin{equation}
\label{steady}
c_\infty(r)=c_0\left[1-\left(\frac{R}{r}\right)^{d-2}\right].
\end{equation}
We now estimate $r_{\rm min}$ from the extreme-value criterion  \cite{rb}  
\begin{equation}
\label{evc}
\int_R^{r_{\rm min}} c(r,t) 2\pi r\,dr=1\,,
\end{equation}
which for the density profile \eqref{steady} becomes 
\begin{eqnarray*}
1=\int_R^{r_{\rm min}} c_0\left[1-\left(\frac{R}{r}\right)^{d-2}\right]\,
2\pi r\,dr.
\end{eqnarray*}
Computing the integral we  find that the dimensionless minimal distance 
$u_{\rm min}=r_{\rm min}/R$ obeys
\begin{eqnarray*}
(\pi c_0 R^2)^{-1}=u_{\rm min}^2-1-
\frac{u_{\rm min}^{2+2\lambda}-1}{1+\lambda}\,.
\end{eqnarray*}

For source flows, the minimal distance grows with time because particles are
driven away from the origin.  In using the connection to pure diffusion in a
space of dimension $d_{\rm eff}$, we note that $d_{\rm eff}<2$ and therefore
we can set the radius of the trap to zero because a diffusing particle is
recurrent and will hit a point trap with certainty if $d_{\rm eff}<2$.  This
equivalent problem has the advantage of being tractable analytically by
elementary means.  Indeed, we look for a scaling solution of the form
\begin{equation}
\label{scal}
c(r,t)=c_0 F(\eta), \quad \eta=\frac{r^2}{4Dt}\,,
\end{equation}
that also satisfies
\begin{equation}
\label{inbound}
c|_{r=0}=0, \quad
c|_{t=0}=c_0.
\end{equation}
By substituting the scaling ansatz \eqref{scal} into the governing diffusion
equation \eqref{diff} we find
\begin{equation}
\label{FF}
\eta F''+(1-\lambda+\eta)F'=0,
\end{equation}
whose solution is
\begin{equation}
\label{FFeta}
F(\eta)=\int_0^\eta d\xi\,\frac{\xi^{\lambda-1}\,e^{-\xi}}{\Gamma(\lambda)}\,.
\end{equation}
Using Eq.~\eqref{evc} together with \eqref{FFeta}, the minimal value
$\eta_{\rm min}=r^2_{\rm min}/4Dt$ of the scaling variable obeys
\begin{eqnarray*}
\frac{\Gamma(\lambda)}{4\pi Dt c_0}=\int_0^{\eta_{\rm min}} 
d\xi\,\xi^{\lambda-1}(\eta_{\rm min}-\xi)\,e^{-\xi}.
\end{eqnarray*}
In the long-time limit, $\eta_{\rm min}\to 0$ and hence the above integral
simplifies to $\frac{\Gamma(\lambda)\Gamma(2)}{\Gamma(\lambda+2)}\,
\eta_{\rm min}^{\lambda+1}$. Thus  we arrive at
\begin{equation}
\label{rmin}
r_{\rm min}\to \left[\frac{\Gamma(\lambda+2)}{\pi c_0}\,
(4Dt)^\lambda\right]^{1/(2\lambda+2)}.
\end{equation}

Let us now illustrate our results by considering some specific examples.  For
sink flow with $\lambda=-1/2$, the effective dimension $d_{\rm eff}=3$ and
\begin{equation*}
r_{\rm min}=R+(\pi c_0)^{-1/2}\,.
\end{equation*}
For source flow with $\lambda=1/2$, $d_{\rm eff}=1$ and
\begin{equation*}
r_{\rm min}\to \left(\frac{9}{4\pi}\,\frac{Dt}{c_0^2}\right)^{1/6}\,.
\end{equation*}
For source flow with $\lambda=1$, $d_{\rm eff}=0$ and
\begin{equation*}
 r_{\rm min}\to \left(\frac{8}{\pi}\,\frac{Dt}{c_0}\right)^{1/4}\,.
\end{equation*}
Note that for true one-dimensional diffusion, the distance from the trap to
the closest surviving particle also scales as $(Dt/c_0^2)^{1/4}$ \cite{rb,WHK,HLKW}.
Finally for very strong source flows with $\lambda\gg 1$, the effective dimension is
$d\ll 0$ and
\begin{equation*}
 r_{\rm min}\to \sqrt{4Dt\,\frac{\lambda}{e}}=\sqrt{\frac{Qt}{\pi e}}\,.
\end{equation*}
In this case, the particles are driven away from the trap in an essentially
deterministic manner and the above scaling dependence of $r_{\rm min}$
follows directly from $\dot r = v \sim Q/r$.
 
The case of pure diffusion in two dimensions is most subtle: the density
profile remains time-dependent, but we cannot seek a solution in a scaling
form because the size of the trap cannot be ignored in two dimensions
\cite{fpp}.  The divergence of the minimal distance with time is also
particularly delicate in this case, namely \cite{rb,HLKW},
\begin{equation}
\label{log}
r_{\rm min}\sim \sqrt{\frac{\ln(Dt/R^2)}{2\pi c_0 \ln[\ln (Dt/R^2)]}}\,.
\end{equation}

\section{Interacting Particle Systems With a Localized Source}
\label{coag}

Consider now the influence of radial flow on an interacting particle systems.
Particularly simple models of this type are irreversible annihilation or
irreversible coalescence \cite{crl,HRS,k2,k3}.  Suppose that the system is
initially empty and that a localized particle source of strength $J$ at the
origin is turned on at time $t=0$.  Particles are emitted at the origin and
are advected by the radial flow field.  The particles also diffuse and react
upon colliding with each other (we set the particle radii equal to one).
These processes may be described by the rate equation
\begin{equation}
\label{rate}
\frac{\partial c}{\partial t}
=D\left(\frac{\partial^2 c}{\partial r^2}
+\frac{d_{\rm eff} -1}{r}\,\frac{\partial c}{\partial r}\right)
-\frac{Dc^2}{\ln(1/c)}
+J\delta({\bf r}).
\end{equation}
Here we employ a phenomenologically-motivated form for the reaction term that
reflects the true two-dimensional spatial nature of the problem.  In the
homogeneous source-free case Eq.~(\ref{rate}) reduces to $\dot
c=-Dc^2/\ln(1/c)$, which leads to $c(t)\sim (Dt)^{-1} \ln(Dt)$ asymptotic
behavior that was rigorously proved to be correct \cite{proof}.

We now employ the same line of reasoning as that given in Refs.~\cite{crl,k3}
to determine the particle density.  When the effective dimension sufficiently
large, $d>4$, particles do not ``see'' each other; {\it i.e.}, they interact
sufficiently weakly that far from the source a non-interacting density
profile, $c(r)\sim r^{-(d-2)}$, should arise.  For $d<4$, the reaction and
diffusion terms balance each other, leading to $c(r)\sim r^{-2}\ln r$.  The
borderline case of $d=4$ needs to be analyzed separately and after some
algebra [see Eqs.~\eqref{c-eq}--\eqref{c-sol} below] it is possible to find a
nested logarithmic correction to the basic $r^{-2}$ behavior: $c(r)\sim
r^{-2}[\ln (\ln r)]^{-1}$.

Having determined the density profile, we can now probe the temporal behavior
of the total number of particles.  Using the fact that the flow field decays
as $1/r$ and that $\dot r= v$, the maximum distance traveled by particles
grows as $\sqrt{t}$.  Thus the total number of particle in the system scales
as $N(t)\sim \int_0^{\sqrt{t}} dr\,r\,c(r)$.  Using this relation, we arrive
at the following conclusions:
\begin{itemize}
\item For sink flows with $\lambda<-1$, the effective dimension obeys $d_{\rm
    eff}>4$.  The density profile and the total number of particles are
      \begin{equation}
       c(r)\sim r^{-(d-2)}=r^{2\lambda}, \quad
       N(t)\sim 1.
      \end{equation}
The sink is thus strong enough to keep the total number of particles finite.

\item For the case of sink flow with $\lambda=-1$, the effective
      dimension is critical, $d_{\rm eff}=4$. The density profile and the
      total number of particles are
      \begin{equation}
       c(r)\sim r^{-2}[\ln (\ln r)]^{-1}, \quad
       N(t)\sim \frac{\ln t}{\ln (\ln t)}.
      \end{equation}
      The total number of particles therefore exhibits an unusually slow
      growth in this marginal regime.

\item For sink flows with $-1<\lambda<0$ and also for source flows, the
      effective dimension obeys $d_{\rm eff}<4$. The density profile and the
      total number of particles become universal
      \begin{equation}
        \label{source}
       c(r)\sim r^{-2}\ln r, \quad
       N(t)\sim (\ln t)^2.
      \end{equation}
\end{itemize}
We conclude that for the case $\lambda\geq -1$ the total number of particles
grows extremely slowly with time, while for $\lambda<-1$ there are just a few
particles in the system.

The results for source flow are seemingly peculiar; here one might naively
expect that the total density grows linearly with time because particles are
driven away from each other by the flow and thus would not interact.  Let us
therefore re-derive some of our results without exploiting the mapping
between radial flow in two dimensions to pure diffusion in variable spatial
dimension.  Consider, for simplicity, the case of strong source flow, {\it
  i.e.}, the P\'eclet number is large.  We may then drop the diffusion terms
and keep just the dominant convection term.  Using this simplification and
assuming that the system is in the steady-state regime, the governing
equation \eqref{rate} becomes
\begin{equation}
\label{strong}
\frac{2\lambda}{r}\,\frac{dc}{dr}=-\frac{c^2}{\ln(1/c)}\,.
\end{equation}
Integrating this equation we obtain
\begin{equation}
\label{sol}
\frac{\ln(1/c)}{c}\simeq A+\frac{r^2}{4\lambda},
\end{equation}
where we use the shorthand notation $A=c_0^{-1}\ln(1/c_0)$.  We estimate the
density $c_0$ near the origin from the conservation law $2\pi
rc_0\,\frac{Q}{2\pi r}=J$ to give $c_0=J/Q$ and therefore
\begin{equation*}
A=\frac{Q}{J}\,\ln\left(\frac{Q}{J}\right). 
\end{equation*}
{}From Eq.~(\ref{sol}), we finally obtain the steady state density profile
\begin{equation}
\label{sollong}
c(r)\simeq \frac{\ln\left[A+\frac{r^2}{4\lambda}\right]}{A+\frac{r^2}{4\lambda}}.
\end{equation}
Sufficiently far from the origin, $r\gg \sqrt{\lambda A}$,
Eq.~(\ref{sollong}) simplifies to $c(r)\sim r^{-2}\ln r$ in agreement with our
previous findings in \eqref{source}.

To determine the total number of particles we note that the steady state
solution formally holds as long as $r\ll \sqrt{Qt/\pi}$.  Using the fact that
the leading edge of the particles advance as $r=\sqrt{Q t/\pi}$, the total
number of particles density then scales as
\begin{displaymath}
N(t)\sim \int_0^{\sqrt{Q t/\pi}} c(r)\,2\pi r\,dr.  
\end{displaymath}
{}From (\ref{sollong}) we then obtain
\begin{equation}
\label{Nt}
N(t)\sim 2\pi\lambda\left\{\ln^2(Dt+A)-\ln^2 A\right\}.
\end{equation}
In the long time limit, Eq.~(\ref{Nt}) qualitatively agrees with our previous
result, $N(t)\sim (\ln t)^2$.  Note, however, that logarithmic behavior
arises only after a short-time linear regime.  For $Dt\ll A$, we get
$N(t)\sim 4\pi\lambda\,\frac{\ln A}{A}\,Dt$, which simplifies to
$N(t)\sim Jt$.  Thus indeed the total density initially grows linearly in
time in accordance with intuition.  The crossover between these two regimes
occurs at $t_c=A/D\sim (\lambda/J)\ln(Q/J)$.

The previous analysis equally applies to annihilation and coalescence.  More
generally, we may also consider irreversible mass-conserving aggregation where
the reaction of a cluster of mass $i$ with a cluster of mass $j$ leads to a
cluster of mass $k=i+j$.  For aggregation, the fundamental quantities are the
densities of various particle species.  Let us set the mass of particles that
are emitted at the origin to unity.  Denote the density of these monomer
particles as $c_1$.  Generally let $c_k$ be the density of particles composed
of $k$ monomers.  We shall seek only the stationary densities $c_k(r)$.
These densities satisfy a modified Smoluchowski equation
\begin{eqnarray}
\label{smol}
\left(\frac{d^2 }{d r^2}+\frac{d_{\rm eff} -1}{r}\,\frac{d }{d r}\right)c_k
&+&\frac{1}{\ln(1/c)}\left(\sum_{i+j=k}c_i c_j - 2c_k c\right)\nonumber\\
&=&-\frac{J}{D}\delta_{k,1}\delta({\bf r}).
\end{eqnarray}
Here we again employed a phenomenologically-motivated form for the reaction
term (see {\it e.g.}, \cite{Meakin}) and we additionally assumed that the
coalescence rate is independent of the masses of the two reactants. (The
latter assumption was originally made by Smoluchowski; it greatly simplifies
the analysis of the infinite system of rate equations \cite{C} .
Furthermore, for diffusion-controlled reactions in two dimensions, the
reaction rate depends on the reactant radii in a weak logarithmic fashion
\cite{ov,fpp}.)

The asymptotic behavior is especially interesting in the marginal case of
$\lambda=-1$ when the effective dimension is critical, $d_{\rm eff}=4$.  Then
the total density satisfies
\begin{equation}
\label{c-eq}
\frac{d^2 c}{d r^2}+\frac{3}{r}\,\frac{d c}{d r}=\frac{c^2}{\ln(1/c)},
\end{equation}
whose leading asymptotic behavior is
\begin{equation}
\label{c-sol}
c=\frac{4}{r^2}\,\frac{1}{\ln(\ln r)}.
\end{equation}
The density of monomers then satisfies 
\begin{eqnarray*}
\frac{d^2 c_1}{d r^2}+\frac{3}{r}\,\frac{d c_1}{d r}&=&\frac{2c}{\ln(1/c)}\,c_1\\
&=&\frac{4}{r^2(\ln r)\ln(\ln r)}\,c_1,
\end{eqnarray*}
whose solution is
\begin{equation}
\label{c1-sol}
c_1\sim \frac{1}{r^2}\,\frac{1}{[\ln(\ln r)]^2}.
\end{equation}
Solving for the first few $k$-mer densities one by one, we observe that they
all have the form
\begin{equation}
\label{ck-seek}
c_k=\frac{1}{r^2}\,F_k(\rho), \quad \rho=\frac{1}{4}\,\ln(\ln r).
\end{equation}
This approach also gives $F_k\sim \rho^{-2}$ for small $k$, but it is better
to derive this result in general rather than guessing this behavior based on
the small-$k$ behavior only.  Thus substituting the ansatz \eqref{ck-seek}
into \eqref{smol} and keeping only the dominant terms we obtain
\begin{equation*}
\frac{d F_k}{d \rho}=\sum_{i+j=k}F_i F_j-\frac{2}{\rho}\,F_k,
\end{equation*}
which is a long-time limit of a system of equations originally solved by
Smoluchowski, see \cite{C}, with $\rho$ playing the role of time.  The
solution admits the scaling form, $F_k=\rho^{-2}e^{-k/\rho}$, and therefore
\begin{equation}
\label{ck-sol}
c_k=\frac{16}{r^2}\,\frac{1}{[\ln(\ln r)]^2}\,\exp\left\{-\frac{4k}{\ln(\ln r)}\right\}.
\end{equation}
Using this result we can verify that the total particle density
$c(r)=\sum_{k\geq 1}c_k(r)$ is consistent with our previous prediction
\eqref{c-sol}; we can also compute the mass density $m(r)=\sum_{k\geq 1}k
c_k(r)=r^{-2}$ and estimate the total mass to be
\begin{equation}
\label{Mt}
M(t)=\int_0^{\sqrt{t}} m(r)\,2\pi r dr=\pi \ln t.
\end{equation}

When $\lambda>-1$, that is, the effective dimension is smaller than critical,
$d_{\rm eff}<4$, the particles do not undergo a sufficient number of
collisions to develop a scaling form like \eqref{ck-sol}.  
Nevertheless, at least the monomer density always exhibits interesting
dynamical behavior.  Indeed, using \eqref{smol} and $c=8(1+\lambda)r^{-2}\ln
r$, which is the precise form of \eqref{source} that includes the correct
amplitude, we find that the monomer density satisfies
\begin{equation}
\label{c1-source}
\frac{d^2 c_1}{dr^2}+\frac{1-2\lambda}{r}\frac{dc_1}{dr}=
\frac{8(1+\lambda)}{r^2}\,c_1\,.
\end{equation}
Solving \eqref{c1-source} gives
\begin{equation}
\label{c1-source-sol}
c_1\sim r^{-n}\,,\quad n=-\lambda+\sqrt{\lambda^2+8(1+\lambda)},
\end{equation}
so that the monomer density has a pure algebraic tail with a
$\lambda$-dependent exponent.

\section{Concluding Remarks}
\label{summ}

Allowing the spatial dimension $d$ to be a free parameter and then developing
theoretical approaches based on this parameter (an expansion about a critical
dimension, dimensional regularization, etc.) has proven extremely fruitful in
field theory and statistical physics \cite{W,i}.  Thus in many treatments of
critical phenomena, non-integer dimensions arise naturally, but this
construction is generally used as an intermediate step toward the ultimate
goal of obtaining information in physically relevant spatial dimensions, such
as $d=1,2,3$.  However, certain complex problems can be mapped onto simpler
ones that are defined in spaces of a non-integer spatial dimension.  We
demonstrated that such a connection arises for a class of two-dimensional
convection-diffusion problems in which the flow field is radial and
proportional to $1/r$.  It would be exciting to realize these
convection-diffusion flows experimentally and hence probe dynamical processes
in spaces whose dimension is not necessarily an integer.

\acknowledgments{We acknowledge financial support from NSF grant CHE0532969
  (PLK) and NSF grant DMR0535503 (SR).}


\begin{thebibliography}{99}

\bibitem{ov}
   A.~A.~Ovchinnikov, S.~F.~Timashev, and A.~A.~Belyi,
   {\it Kinetics of Diffusion Controlled Chemical Processes} (Nova  Science,1989).
   
\bibitem{redner} 
    S.~Redner and F.~Leyvraz, in {\it Fractals and Disordered Systems}, 
    Vol. II eds. A. Bunde and S. Havlin (Springer-Verlag, Berlin, 1993).
    
\bibitem{oshanin}
   G.~Oshanin, M.~Moreau, and S.~Burlatsky, Adv.\ Coll.\ Inter.\ Sci. 
   {\bf 49}, 1 (1994).  

\bibitem{priv}
   {\it Non-equilibrium Statistical Mechanics in One Dimension}, ed.\  V.~Privman
   (Cambridge University Press, New York, 1997).

\bibitem{dani}
   D.~ben-Avraham and S.~Havlin,
   {\it Diffusion and Reactions in Fractals and Disordered Systems}
   (Cambridge University Press, New York, 2000).
    
\bibitem{fpp} S. Redner, {\it A Guide to First-Passage Processes} (Cambridge
  University Press, New York, 2001).

\bibitem{rb} 
     S.~Redner and D.~ben-Avraham, J. Phys.\ A {\bf 23}, L1169 (1990).

\bibitem{k} 
     P.~L.~Krapivsky, Phys. Rev. E {\bf 47}, 1199 (1993).

\bibitem{WHK} G. H. Weiss, S. Havlin, and R. Kopelman, Phys.\ Rev.\ A {\bf
    39}, 466 (1989).

\bibitem{HLKW} S. Havlin, H. Larralde, R. Kopelman, and G. H. Weiss, Physica A
  {\bf 169}, 337 (1990).

\bibitem{crl} 
     Z.~Cheng, S.~Redner, and F.~Leyvraz, Phys.\ Rev.\ Lett.\ {\bf 62},
     2321 (1989).

\bibitem{HRS} H. Hinrichsen, V. Rittenberg, and H. Simon, J. Stat.\ Phys.\
  {\bf 86}, 1203 (1997).

\bibitem{k2} 
   P.~L.~Krapivsky, Physica A {\bf 198}, 157 (1993).
  
\bibitem{k3} 
   P.~L.~Krapivsky, Phys.\ Rev.\ E {\bf 49}, 3233 (1994).
  
\bibitem{proof} 
     M.~Bramson and D.~Griffeath, Z.\ Wahrsch.\ Verw.\ Gebiete {\bf 53},
     183 (1980).

\bibitem{Meakin}
     P.~Meakin, Physica A {\bf 165}, 1 (1990). 
     
\bibitem{C} 
    S. Chandrasekhar, Rev.\ Mod.\ Phys.\ {\bf 15}, 1 (1943).

\bibitem{W}
     K. G. Wilson and M. E. Fisher, Phys.\ Rev.\ Lett.\ {\bf 28}, 240
    (1972); K. G. Wilson, Phys.\ Rev.\ Lett.\ {\bf 28}, 548 (1972).

\bibitem{i} 
     C.~Itzykson and J.-M.~Drouffe, {\it Statistical Field Theory}
     (Cambridge University Press, Cambridge, 1989).


\end{thebibliography}
\end{document}